\documentclass[11pt]{article}
\usepackage{amsmath}
\usepackage{amssymb}
\usepackage{amsthm}
\usepackage{xcolor}
\usepackage{caption}
\usepackage{calc}
\usepackage{hyperref}
\usepackage{graphicx}
\usepackage{wrapfig}
\usepackage{enumitem}
\usepackage{cite}
\usepackage{geometry}
\usepackage{algorithm}
\usepackage[noend]{algpseudocode}
\newcounter{algsubstate}
\renewcommand{\thealgsubstate}{\alph{algsubstate}}

\begin{document}

\title{Denoising deterministic networks using iterative Fourier transforms}

 \author{H. Robert Frost$^{1}$}
\date{}
\maketitle
\begin{center}
\textit{
$^1$Department of Biomedical Data Science\\
Dartmouth College \\
Hanover, NH 03755, USA \\
rob.frost@dartmouth.edu
}
\end{center}

\begin{abstract}
We detail a novel Fourier-based approach (IterativeFT) for identifying deterministic network structure in the presence of both edge pruning and Gaussian noise. This technique involves the iterative execution of forward and inverse 2D discrete Fourier transforms on a target network adjacency matrix. The denoising ability of the method is achieved via the application of a sparsification operation to both the real and frequency domain representations of the adjacency matrix with algorithm convergence achieved when the real domain sparsity pattern stabilizes. To demonstrate the effectiveness of the approach, we apply it to noisy versions of several deterministic models including Kautz, lattice, tree and bipartite networks. For contrast, we also evaluate preferential attachment networks to illustrate the behavior on stochastic graphs. We compare the performance of IterativeFT against simple real domain and frequency domain thresholding, reduced rank reconstruction and locally adaptive network sparsification.  Relative to the comparison network denoising approaches, the proposed IterativeFT method provides the best overall performance for lattice and Kuatz networks with competitive performance on tree and bipartite networks. Importantly, the InterativeFT technique is effective at both filtering noisy edges and recovering true edges that are missing from the observed network.
\end{abstract}

\section{Introduction}\label{sec:intro}

We have recently explored the characteristics of a family of iterative Fourier transformation \cite{Bracewell:2000ua} algorithms that involve the repeated execution of forward and inverse discrete Fourier transformations on vector or matrix valued data (see \cite{frost2026iterativeexecutiondiscreteinverse} for details). Since the sequential execution of a forward and inverse Fourier transform returns the original data, the simple realization of this iteration would act as the identity function irrespective of the number of iterations. However, interesting behavior can be obtained when functions $h()$ and $g()$ are executed on the real domain and frequency domain data respectively. In particular, we found that when $h()$ and $g()$ are thresholding functions (e.g., functions that set all elements whose absolute value or magnitude is less than the mean to 0), the iteration will converge to a stable real domain sparsity pattern. We refer to the sparsification variant of this approach as the IterativeFT method. Importantly, if the input data is a combination of a periodic or structured signal and noise, the method will act as a denoising operation with the output at convergence close to the original signal (see results in \cite{frost2026iterativeexecutiondiscreteinverse}). This technique can be applied to any valid input for a discrete Fourier transformation and we originally evaluated it on both vector and matrix valued inputs for a range of signal and noise combinations.

In this paper, we explore the application of the IterativeFT method to network adjacency matrices for the purpose of extracting the true underlying structure of a noisy and pruned network. In particular, we consider the situation where the observed weighted and directed network containing $n$ nodes is represented by an $n \times n$ adjacency matrix $\mathbf{X}$ that is generated by the addition of Gaussian noise to a randomly pruned version of the true underlying network ($\mathbf{A}$):

\begin{equation}\label{eqn:noisy_network}
\mathbf{X} = \mathbf{A} * \mathbf{P} + \mathbf{E}
\end{equation}

\noindent where $\mathbf{P}$ is a matrix of binary elements controlling which edges are pruned from the network (i.e., $p_{i,j} = 0$ if an edge from node $i$ to $j$ should be removed and 1 if it should be retained), $*$ is the Hadamard product, and $\mathbf{E}$ is a matrix of random Gaussian noise (i.e., $e_{i,j} \sim \mathcal{N}(0, \epsilon)$). In this scenario, recovery of $\mathbf{A}$ requires both the removal of false edges and detection of missing edges. 

A wide range of techniques have been used for network denoising including both simple or adaptive thresholding of edge weights \cite{Foti:2011aa, PhysRevE.93.012304, Serrano:2009aa}, linear or non-linear low rank approximations (e.g., SVD or autoencoder-based reduced rank reconstruction), and edge prediction techniques \cite{Clauset:2008aa}. While existing techniques can be effective in certain scenarios, most methods share two important limitations: 1) they focus on either noisy edge removal or missing edge prediction but not both, and 2) they require the specification of hyperparameters that control method behavior (e.g., statistical significance for probabilistic methods) or encode assumptions about the structure of the underlying true network (e.g., rank for low rank approximation methods, assumption of a hierarchical random graph model, etc.). These limitations motivated our application of the IterativeFT technique to noisy network adjacency matrices. Importantly, the IterativeFT method does not require hyperparameter tuning or the specification of the true network model and is effective at both removing noisy edges and estimating missing edges. As detailed in the simulation study results in this paper, we find that the IterativeFT method is particularly effective relative to other techniques when the ground truth network follows a deterministic model (e.g., Kautz \cite{Kautz1968} or lattice graphs).

\section{Methods}\label{sec:methods}

\subsection{IterativeFT method}\label{sec:iterft}

Algorithm \eqref{alg:iterft} details the application of the IterativeFT method to a network adjacency matrix. Specifically, the method iteratively applies forward and inverse 2D discrete Fourier transforms to the input network adjacency matrix with interspersed thresholding operations and convergence obtained when the real domain sparsity pattern stabilizes.

\begin{algorithm}
\caption{IterativeFT method for network adjacency matrix denoising}
\label{alg:iterft}
\hspace*{\algorithmicindent} \textbf{Inputs:}
\begin{itemize}
\setlength\itemsep{0em}
\item $\mathbf{X} \in \mathbb{R}^{n \times n}$ \Comment{Adjacency matrix for network with $n$ vertices.}
\item $i_m$ \Comment{The maximum number of iterations}
\end{itemize}
\hspace*{\algorithmicindent} \textbf{Outputs:}
\begin{itemize}
\setlength\itemsep{0em}
\item $\mathbf{Y} \in \mathbb{R}^n$ \Comment{Denoised adjacency matrix}
\item $i_c$ \Comment{Number of iterations completed}
\end{itemize}
\begin{algorithmic}[1]
\State $i = 1$ \Comment{Initialize iteration index}
\State $\mathbf{X}_0 = \mathbf{X}$ \Comment{Initialize $\mathbf{X}_i$}
\While {$i \leq i_m$}
\State $\mathbf{X}^*_i = \mathbf{X}_{i-1}$ 
\State $\mathbf{X}^*_i = threshold(|\mathbf{X}^*_i|, \overline{|\mathbf{X}^*_i|})$ \Comment{Set edge weights whose absolute value is below the mean absolute value to 0}
\If{$sparseMatch(\mathbf{X}^*_i, \mathbf{X}^*_{i-1})$}  
  \State break \Comment{If sparsity pattern is identical, stop the iteration}
\EndIf
\State $\mathbf{W}_i = \textit{dft}(\mathbf{X}^*_i)$ \Comment{Compute 2D discrete Fourier transform of $\mathbf{X}^*_i$}
\State $\mathbf{W}^*_i = threshold(|\mathbf{W}_i|, \overline{|\mathbf{W}_i|})$  \Comment{Set complex elements whose magnitude is below the mean magnitude to 0}
\State $\mathbf{X}_{i} = \textit{dft}^{-1}(\mathbf{W}^*_i)$ \Comment{Compute inverse 2D discrete Fourier transform of $\mathbf{W}^*_i$}
\State $i=i+1$ \Comment{Increment iteration index}
\EndWhile
\Return $(\mathbf{X}^*_i,i)$ \Comment{Return the final thresholded adjacency matrix}
\end{algorithmic}
\end{algorithm}

Iterating between real domain and frequency domain sparsification is motivated by the discrete Fourier transform uncertainty principal \cite{doi:10.1137/0149053}, which constrains the total number of zero values in the real and frequency domains. Specifically, sparsity in one domain will reduce sparsity in the other domain and thereby lead to convergence at a non-trivial solution, ie., a matrix that doesn't have all 0 values. Both general and sparsification versions of Algorithm \ref{alg:iterft} are supported by the IterativeFT R package available at
\href{https://hrfrost.host.dartmouth.edu/IterativeFT}{https://hrfrost.host.dartmouth.edu/IterativeFT}.

\subsection{Comparison methods}\label{sec:comp_methods}

We compared the network denoising performance of the IterativeFT method against four other approaches:

\begin{itemize}
\item \textbf{Real domain thresholding}: This method performs a simple thresholding operation on $\mathbf{X}$ where all elements whose absolute value is lower than the mean absolute value are set to 0.
\item \textbf{Frequency domain thresholding}: This method performs a simple frequency domain thresholding operation, i.e., the denoised adjacency matrix is generated as $dft^{-1}(threshold(dft(\mathbf{X})))$. Similar to the real domain thresholding, all complex values whose magnitude is lower than the mean magnitude are set to 0.
\item \textbf{Low rank reconstruction}: This technique computes the singular value decomposition (SVD) of the adjacency matrix, $\mathbf{X} = \mathbf{U} \boldsymbol{\Sigma} \mathbf{V}^T$, and uses that to create a rank 3 reconstruction as $\mathbf{X}_{lr} = \mathbf{U}[,1{:}3] \boldsymbol{\Sigma}[1{:}3,1{:}3] \mathbf{V}[,1{:}3]^T$.
\item \textbf{Locally adaptive network sparsification (LANS)}: We used the LANS method \cite{Foti:2011aa} as implemented by the \textit{backbone\_from\_weighted(X, alpha=0.05, mtc="BH")} function in v3.0.3 of the R \textit{backbone} package \cite{Zachary-P-Neal:2022aa}. 
\end{itemize}

\subsection{Simulated network models}\label{sec:models}

To evaluate the relative performance of the IterativeFT technique, we applied it and the four comparison methods detailed in Section \ref{sec:comp_methods}  to noisy and pruned versions of networks generated according to four deterministic (Kautz, lattice, tree, and full bipartite) and one stochastic model (preferential attachment). Creation of the noisy and pruned networks was realized using the following steps:

\begin{itemize}
\item Create the ground truth network (detailed below for each model).
\item Randomly remove a specific proportion of the edges (default of 0.25).
\item Add iid Gaussian noise with mean 0 and customizable standard deviation (SD) (default of 0.25) to every element of the pruned adjacency matrix.
\end{itemize}

\noindent This process was repeated 10 times for each model across each tested pruning proportion (from 0.05 to 0.95 by 0.05) and noise SD (from 0 to 1 by 0.05). The ground truth networks were created as follows for each model:

\begin{enumerate}
\item \textbf{Kautz}: A Kautz network \cite{Kautz1968} with 108 total vertices was created with an alphabet size of 3 and label size of 3 using the \textit{make\_kautz\_graph()} function from v2.2.1 of the \textit{igraph} R package \cite{Csardi:2006aa}.
\item \textbf{Lattice}: A $10 \times 10$ lattice network (100 total vertices) was created using the \textit{make\_lattice()} \textit{igraph} function.
\item \textbf{Tree}: A 108 vertex undirected tree network with 3 child nodes per non-leaf vertex was created using the \textit{make\_tree()} \textit{igraph} function.
\item \textbf{Full bipartite}: A 108 vertex full bipartite network was created using the \textit{make\_full\_bipartite\_graph()} \textit{igraph} function.
\item \textbf{Preferential attachement}: A random 108 vertex undirected preferential attachment network with attachment power 1 and zero appeal of 1 was simulated using the \textit{sample\_pa()} \textit{igraph} function.
\end{enumerate}

\subsection{Evaluation metrics}\label{sec:metrics}

To evaluate the denoising performance of each method we computed both classification metrics (specifically the F1 score) and the mean squared error (MSE) between the ground truth and denoised adjacency matrices. For computing the F1 score, we assumed that any edge with a non-zero value in the denoised matrix represents a valid edge. Both metrics provide useful complementary insight into method performance. Specifically, the ground truth adjacency matrices for most of the tested models are extremely sparse, which enables very low MSE values for the trivial (and unhelpful) solution of an adjacency matrix with all 0 values. In contrast, important network structure can also be recovered even if the generated adjacency matrix is fully connected (e.g., the results achieved by the low rank approximation method on the full bipartite graph); this performance is captured by the MSE but not by the F1 score.

\section{Results}\label{sec:results}

Each of the sections below detail the relative performance of IterativeFT and the four comparison techniques (real thresholding, frequency thresholding, low rank reconstruction, and LANS) on five different noisy network models (Kautz, lattice, tree, full bipartite and preferential attachment). To clearly visualize the structure of each network model and the denoising performance of the evaluated methods, figures (Figs \ref{fig:kautz_example}, \ref{fig:lattice_example}, \ref{fig:tree_example}, \ref{fig:bipartite_example}, and \ref{fig:pa_example}) are included for a single simulated example that contain eight versions of the associated network adjacency matrix (ground truth, pruned, pruned and noisy, and denoised by each of the five evaluated techniques). To characterize relative performance across a wider range of noise and pruning levels, the mean F1 and MSE values for each denoising method are plotted in Figs \ref{fig:kautz_multiple}, \ref{fig:lattice_multiple}, \ref{fig:tree_multiple}, \ref{fig:bipartite_multiple}, and \ref{fig:pa_multiple}.

\clearpage

\subsection{Kautz network results}\label{sec:kautz_results}

\begin{figure}[h]
\begin{center}
\includegraphics[width=0.9\textwidth]{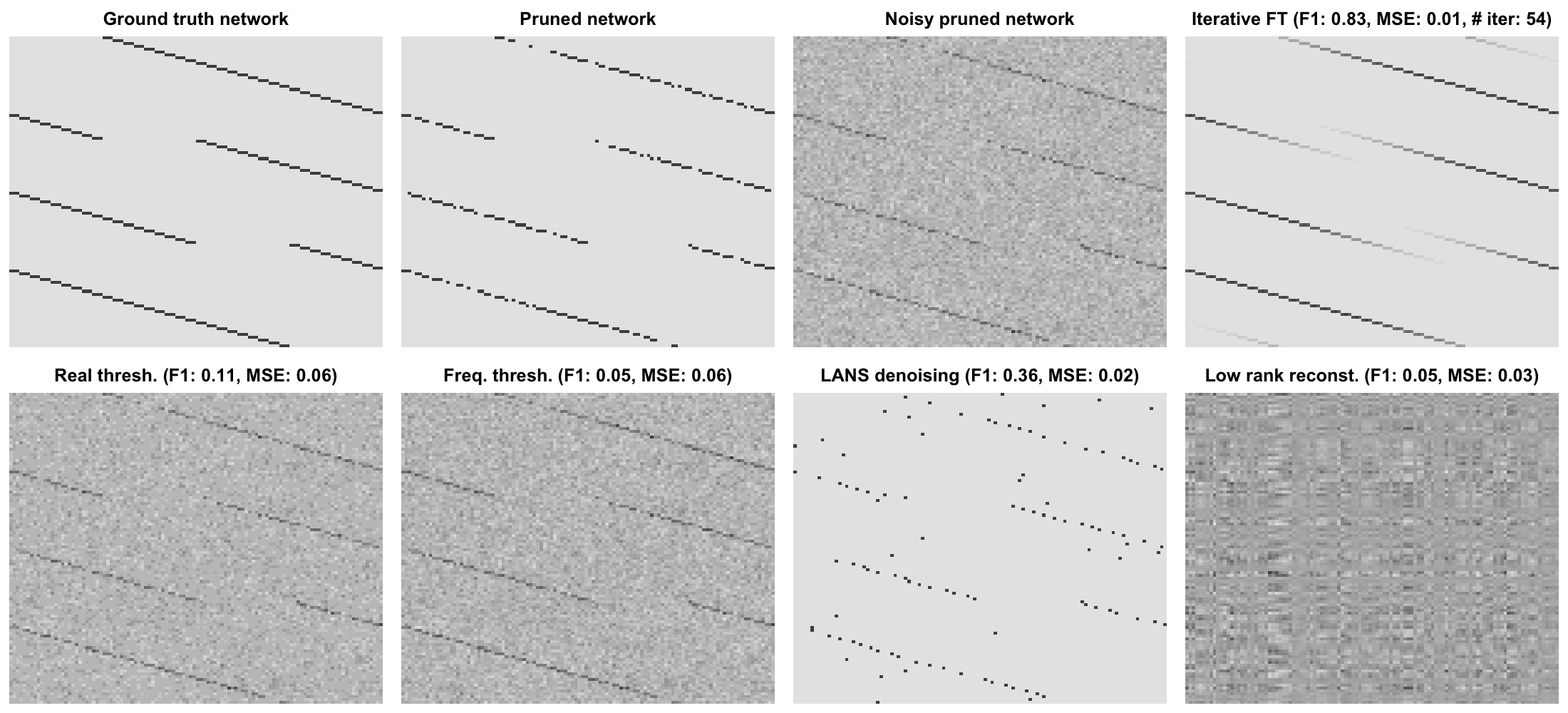}
\end{center}
\caption{Visualization of adjacency matrices associated with the ground truth, pruned, noisy, and denoised versions of a 3,3 Kautz network as detailed in Sections \ref{sec:comp_methods}, \ref{sec:models}, and \ref{sec:metrics}.}
\label{fig:kautz_example}
\end{figure}

\begin{figure}[h]
\begin{center}
\includegraphics[width=0.9\textwidth]{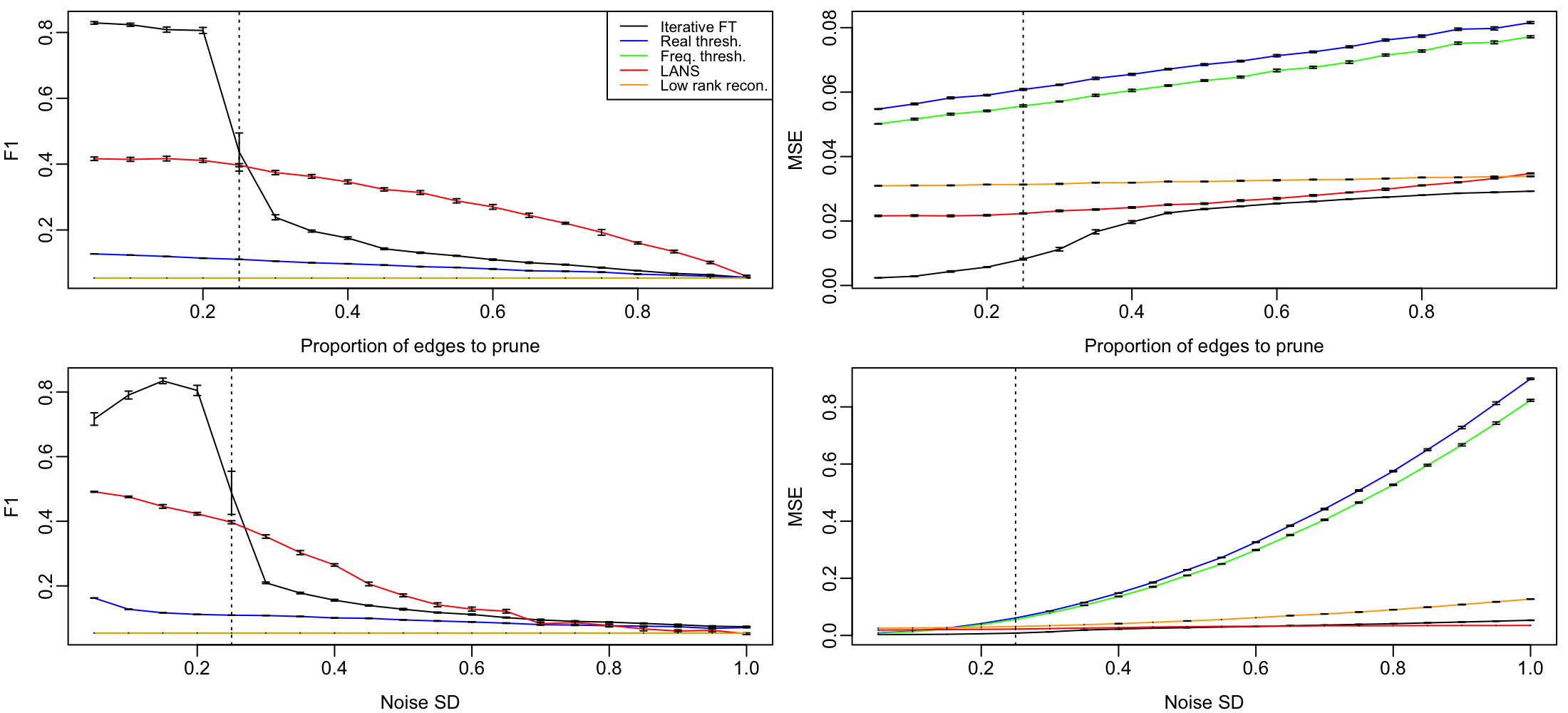}
\end{center}
\caption{Mean F1 score and MSE of evaluated denoising methods for the Kautz model for a range of pruning proportions and noise SD values. The vertical line indicates the default value used for both the example in Fig \ref{fig:kautz_example} and when the other parameter is varied. Error bars represent the standard error of the mean.}
\label{fig:kautz_multiple}
\end{figure}

As shown in \ref{fig:kautz_example} and \ref{fig:kautz_multiple}, the IterativeFT method has the best overall performance on the Kautz model according to both the F1 score and MSE. For higher pruning proportions, the LANS technique provides the best F1.

\clearpage

\subsection{Lattice network results}\label{sec:lattice_results}

\begin{figure}[h]
\begin{center}
\includegraphics[width=0.9\textwidth]{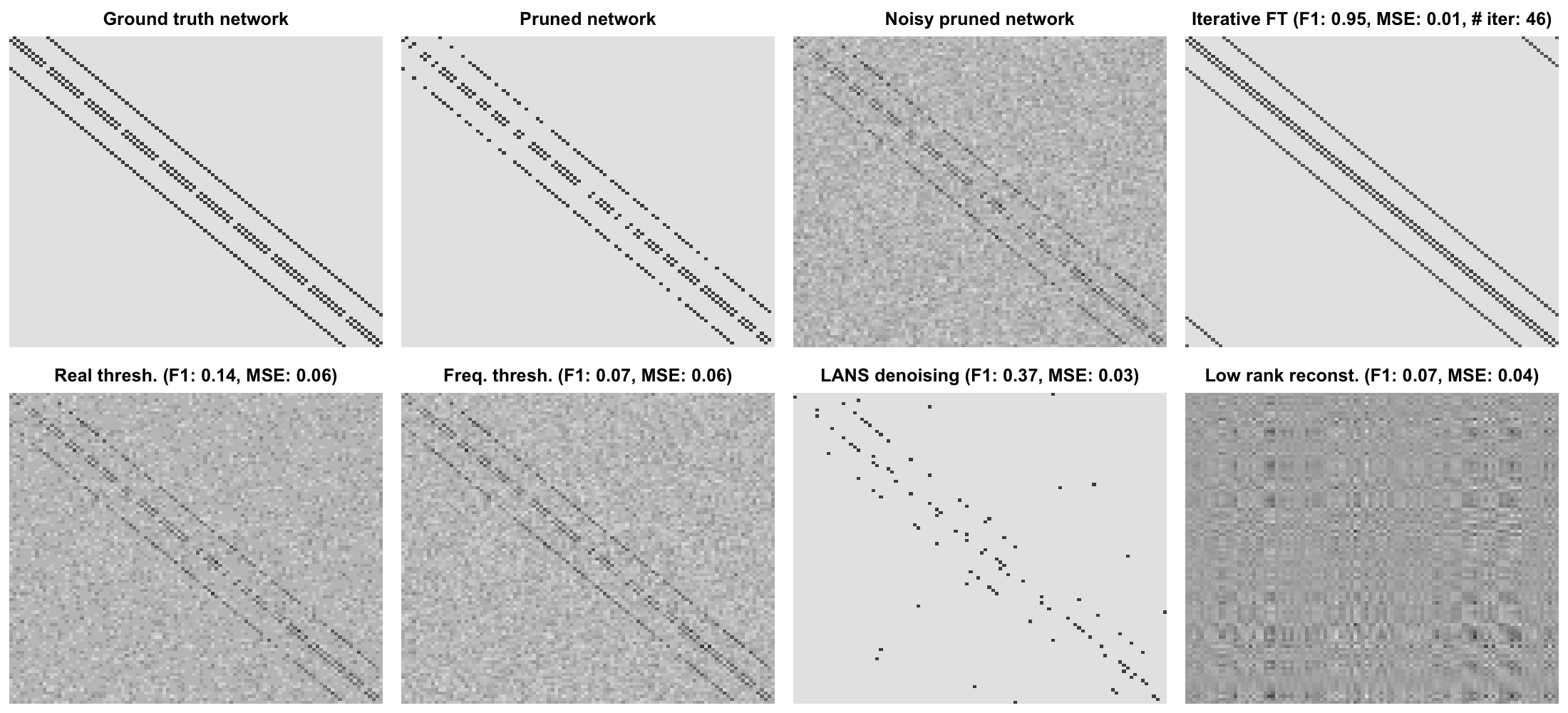}
\end{center}
\caption{Visualization of adjacency matrices associated with the ground truth, pruned, noisy, and denoised versions of a $10 \times 10$ lattice network as detailed in Sections \ref{sec:comp_methods}, \ref{sec:models}, and \ref{sec:metrics}.}
\label{fig:lattice_example}
\end{figure}

\begin{figure}[h]
\begin{center}
\includegraphics[width=0.9\textwidth]{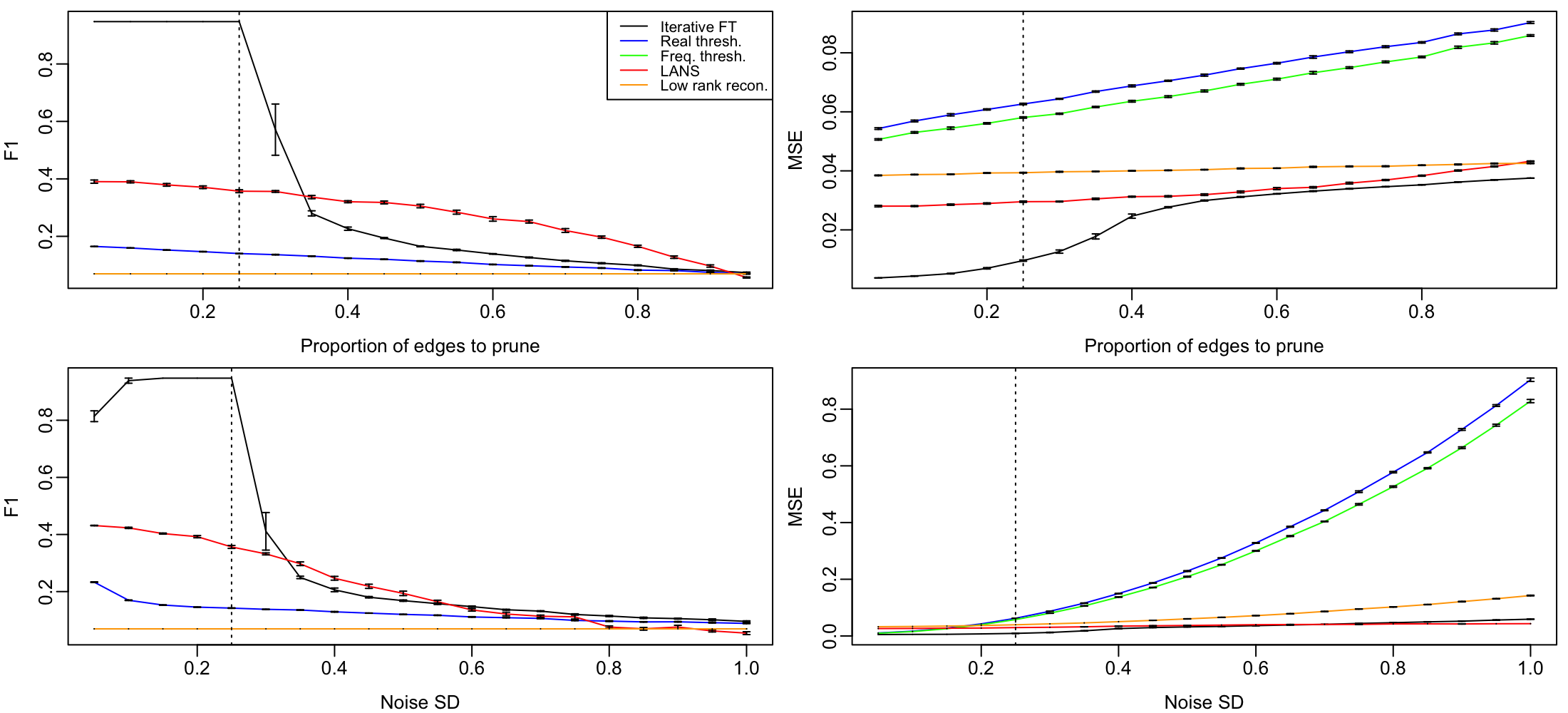}
\end{center}
\caption{Mean F1 score and MSE of evaluated denoising methods for the lattice model for a range of pruning proportions and noise SD values. The vertical line indicates the default value used for both the example in Fig \ref{fig:lattice_example} and when the other parameter is varied. Error bars represent the standard error of the mean.}
\label{fig:lattice_multiple}
\end{figure}

Similar to the Kautz model, the IterativeFT method has the best overall performance on the lattice model according to both the F1 score and MSE as shown in \ref{fig:lattice_example} and \ref{fig:lattice_multiple}. For higher pruning proportions, the LANS technique again provides the best F1.

\clearpage

\subsection{Tree network results}\label{sec:tree_results}

\begin{figure}[h]
\begin{center}
\includegraphics[width=0.9\textwidth]{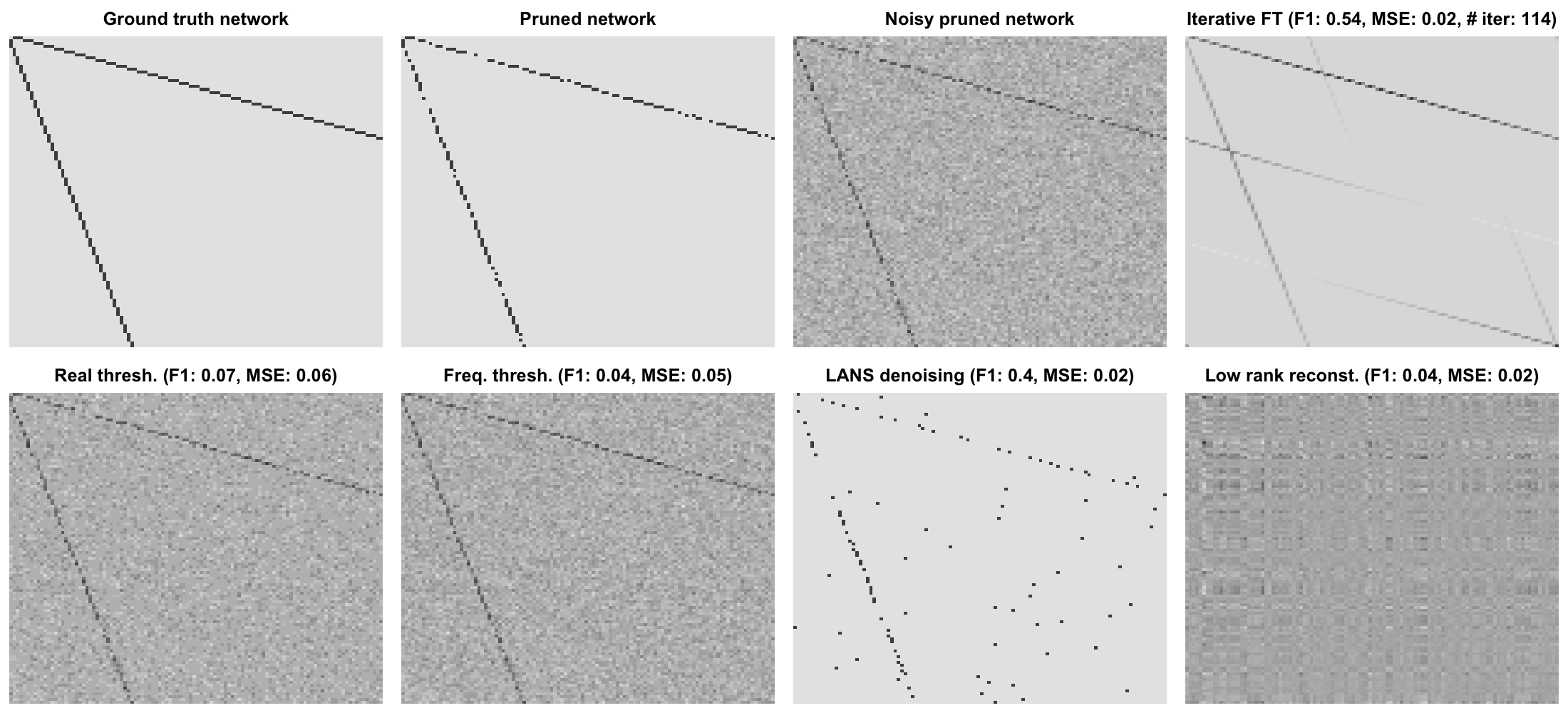}
\end{center}
\caption{Visualization of adjacency matrices associated with the ground truth, pruned, noisy, and denoised versions of a 108 vertex tree network with 3 child nodes per non-leaf vertices as detailed in Sections \ref{sec:comp_methods}, \ref{sec:models}, and \ref{sec:metrics}.}
\label{fig:tree_example}
\end{figure}

\begin{figure}[h]
\begin{center}
\includegraphics[width=0.9\textwidth]{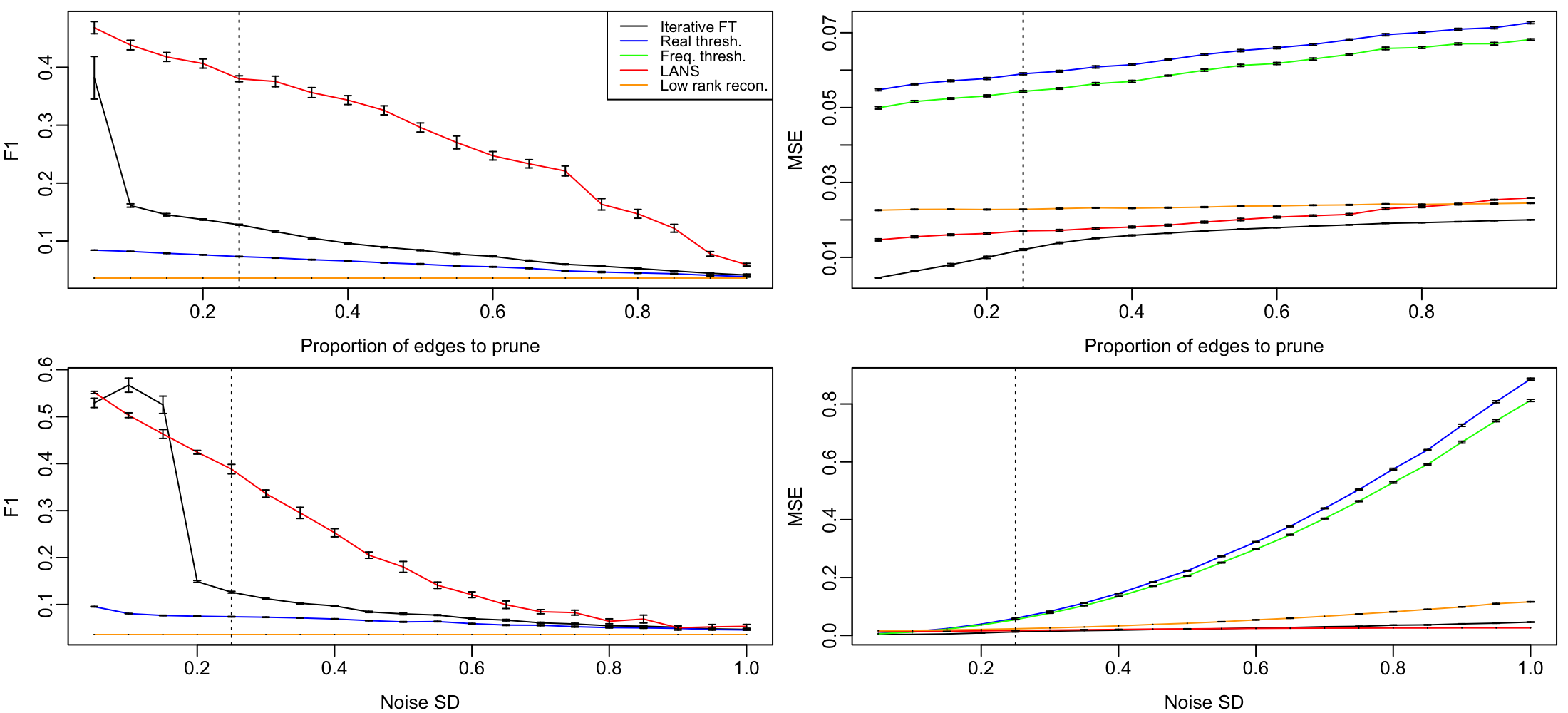}
\end{center}
\caption{Mean F1 score and MSE of evaluated denoising methods for the tree model for a range of pruning proportions and noise SD values. The vertical line indicates the default value used for both the example in Fig \ref{fig:tree_example} and when the other parameter is varied. Error bars represent the standard error of the mean.}
\label{fig:tree_multiple}
\end{figure}

As shown in \ref{fig:tree_example} and \ref{fig:tree_multiple}, the IterativeFT method has the best overall performance on the tree model according to MSE and is second behind LANS according to the F1 score. Similar to the Kautz and lattice results, the performance of LANS relative to IterativeFT improves at higher prunning proportions.

\clearpage

\subsection{Bipartite network results}\label{sec:bipartite_results}

\begin{figure}[h]
\begin{center}
\includegraphics[width=0.9\textwidth]{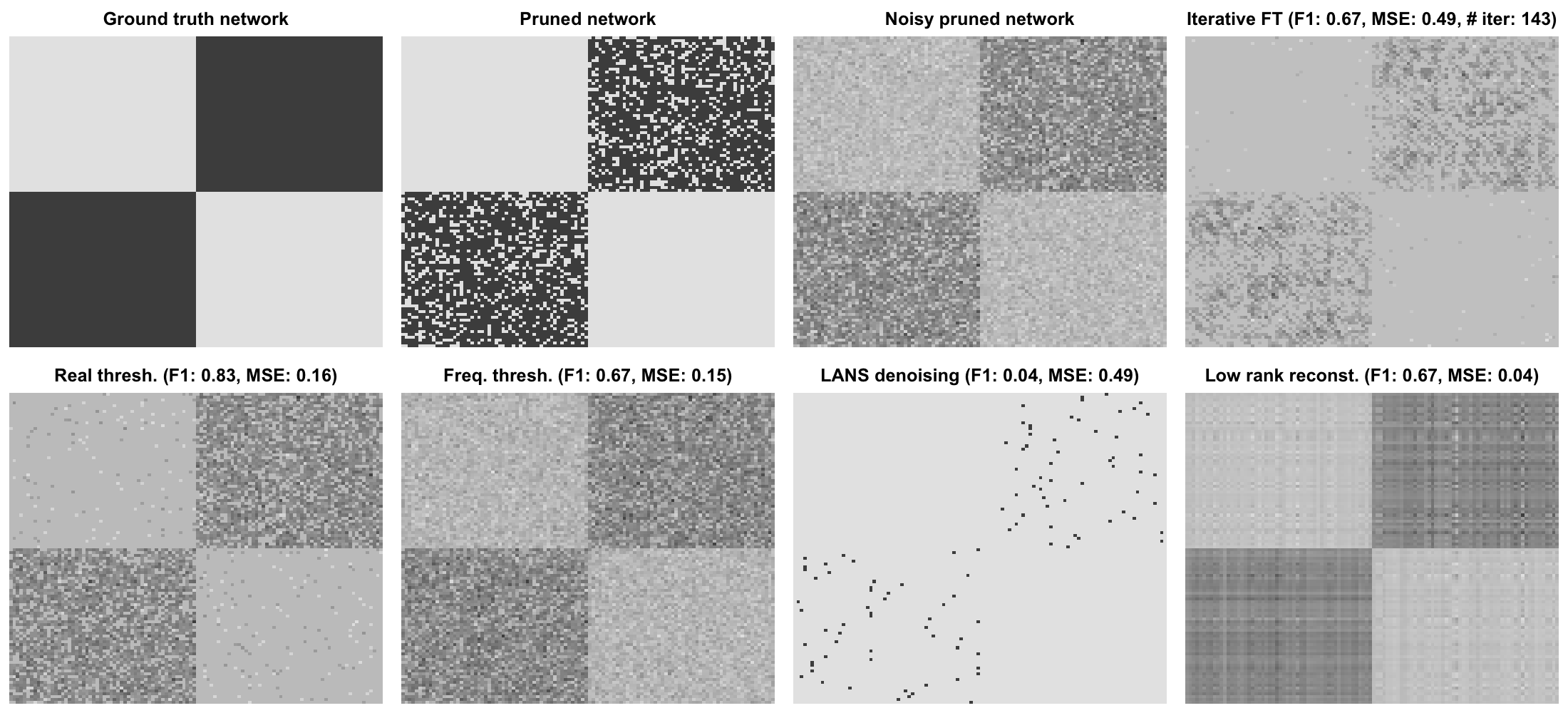}
\end{center}
\caption{Visualization of adjacency matrices associated with the ground truth, pruned, noisy, and denoised versions of a 108 vertex full bipartite network as detailed in Sections \ref{sec:comp_methods}, \ref{sec:models}, and \ref{sec:metrics}.}
\label{fig:bipartite_example}
\end{figure}

\begin{figure}[h]
\begin{center}
\includegraphics[width=0.9\textwidth]{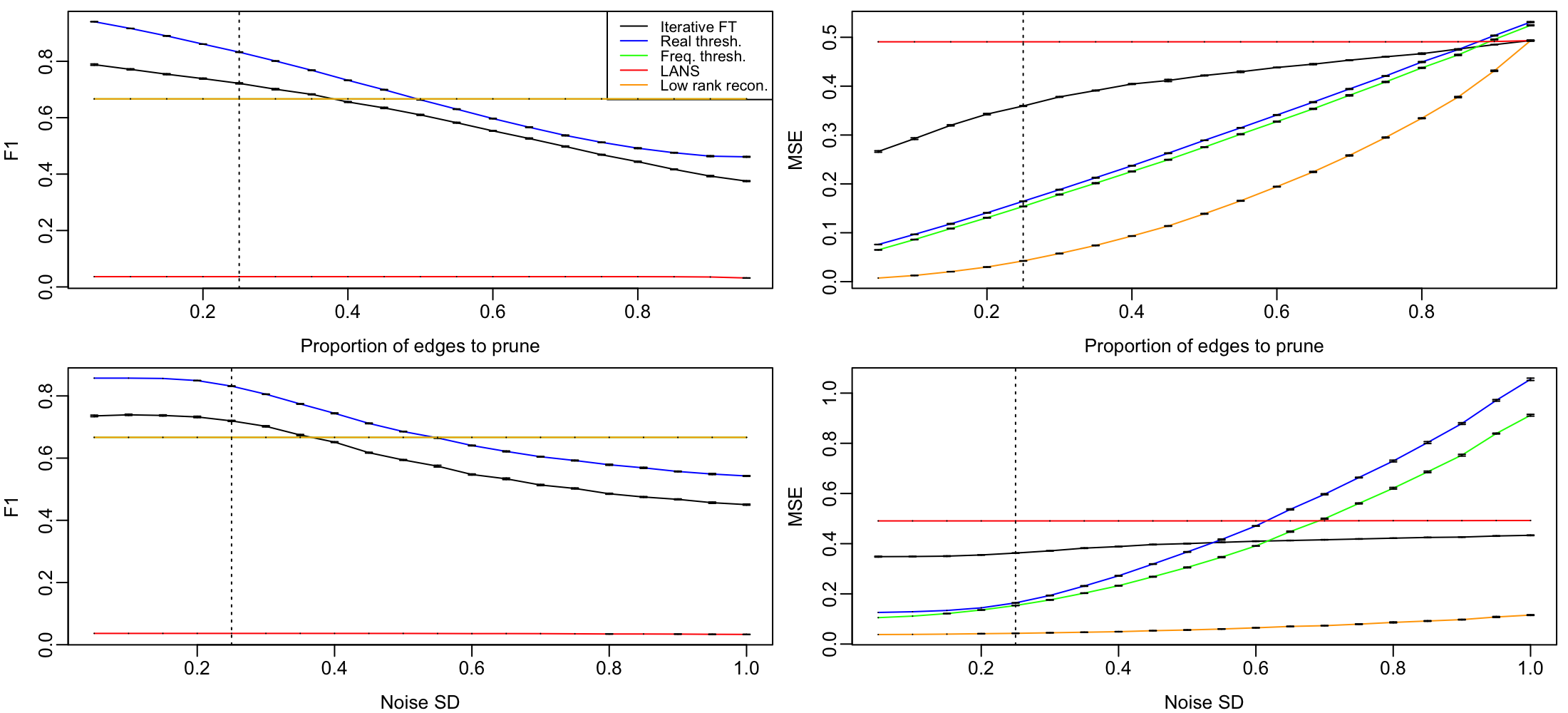}
\end{center}
\caption{Mean F1 score and MSE of evaluated denoising methods for the full bipartite model for a range of pruning proportions and noise SD values. The vertical line indicates the default value used for both the example in Fig \ref{fig:bipartite_example} and when the other parameter is varied. Error bars represent the standard error of the mean.}
\label{fig:bipartite_multiple}
\end{figure}

As shown in \ref{fig:bipartite_example} and \ref{fig:bipartite_multiple}, simple real thresholding is generally superior to the other techniques on the full bipartite model according to the F1 score. By contrast, the low rank reconstruction approach is dominant according to the MSE. The IterativeFT method is competitive according to the F1 score but performs poorly on MSE.

\clearpage

\subsection{Preferential attachement network results}\label{sec:pa_results}

\begin{figure}[h]
\begin{center}
\includegraphics[width=.9\textwidth]{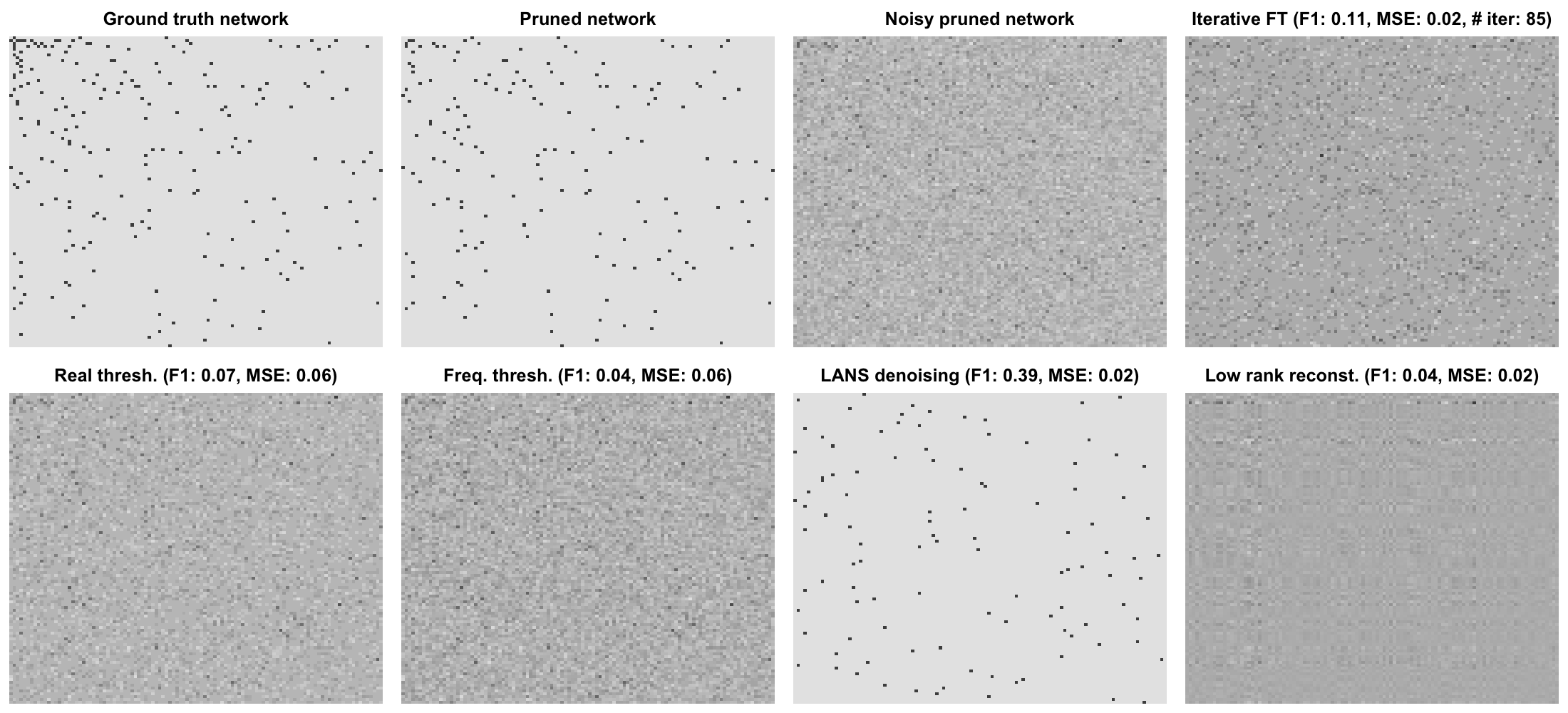}
\end{center}
\caption{Visualization of adjacency matrices associated with the ground truth, pruned, noisy, and denoised versions of a preferential attachement network as detailed in Sections \ref{sec:comp_methods}, \ref{sec:models}, and \ref{sec:metrics}.}
\label{fig:pa_example}
\end{figure}

\begin{figure}[h]
\begin{center}
\includegraphics[width=.9\textwidth]{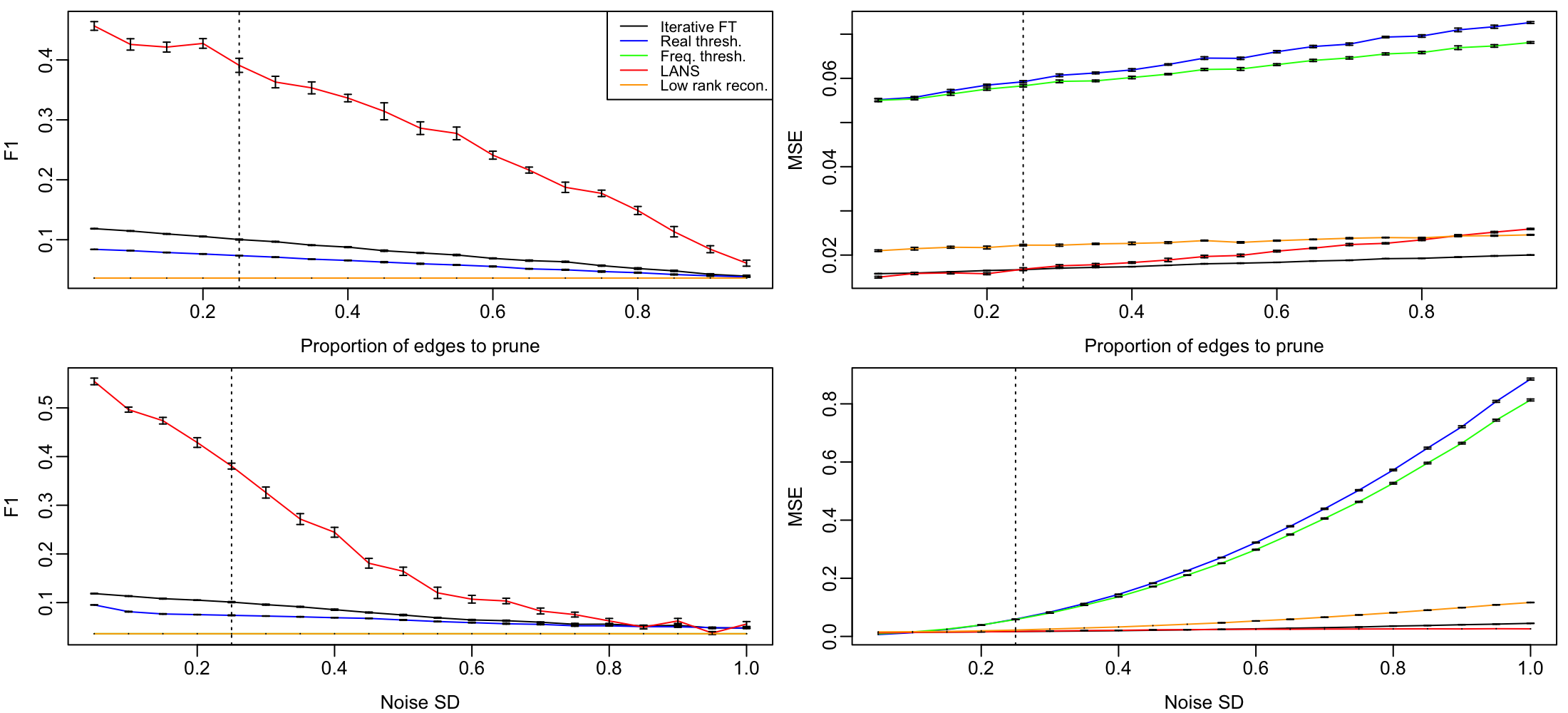}
\end{center}
\caption{Mean F1 score and MSE of evaluated denoising methods for the preferential attachment model for a range of pruning proportions and noise SD values. The vertical line indicates the default value used for both the example in Fig \ref{fig:pa_example} and when the other parameter is varied. Error bars represent the standard error of the mean.}
\label{fig:pa_multiple}
\end{figure}

As shown in \ref{fig:pa_example} and \ref{fig:pa_multiple}, the LANS method has the best overall performance on the preferential attachment model according to the F1 score and near the best according to MSE. As expected, the IterativeFT method is not effective on this network model given the lack of regular structure in the ground truth adjacency matrix.

\clearpage

\section{Discussion and conclusions}\label{sec:conclusion}

In this paper, we explored the network denoising performance of the IterativeFT method, which performs the repeated execution of discrete and inverse discrete 2D Fourier transforms on a network adjacency matrix with interleaved thresholding operations. To evaluate the IterativeFT technique, we compared it against four other denoising techniques (real thresholding, frequency thresholding, low rank reconstruction and locally adaptive network sparsification) on pruned and noisy versions of four deterministic network models (Kautz, tree, lattice and full bipartite) and one stochastic model (preferential attachment). This evaluation revealed that the IterativeFT can effectively recover deterministic network structure in the presence of both edge pruning and Gaussian noise. IterativeFT provided the best overall performance on both the Kautz and lattice models across the range of evaluated pruning proportions and noise levels and provided competitive performance on the tree and bipartite models. In contrast to the results on deterministic models, the IterativeFT method is unable to identify the underlying structure of stochastic networks such as preferential attachment model. 

\subsection{Limitations}

While these initial results are encouraging, there are several important limitations to note:
\begin{itemize}
\item The evaluation tested just five total network models with a fixed network size and fixed model hyperparameters. 
\item The evaluation was limited to four comparison denoising methods and did not explore testing the full range of method parameter values (e.g., only linear low rank reconstruction with a rank of 3, LANS was tested with a fixed significance level, etc.).
\item Most importantly, the IterativeFT method can generate artifactual network structures. This behavior can be seen in Figs \ref{fig:kautz_example}, \ref{fig:lattice_example}, and \ref{fig:tree_example} for the Kautz, tree and lattice examples. For example, IterativeFT introduces false edges in the bottom left and top right corner of the Kautz adjacency matrix along with false extensions of the true edge bands.
\end{itemize}

\subsection{Future directions}

Areas for future work include on this method include:
\begin{itemize}
\item Exploring approaches to mitigate the generation of artifactual network structures, e.g., ensemble methods that combine IterativeFT with sparsification methods such as LANS.
\item Performing a more extensive simulation study that includes a broader range of network models, different network sizes and alternative forms of noise.
\item Evaluating the comparative performance of IterativeFT against a broader range of existing network denoising techniques.
\item Exploring the theoretical basis for the observed denoising performance.
\item Applying the technique to real networks.
\end{itemize}

\section*{Acknowledgments}

This work was funded by National Institutes of Health grants R35GM146586 and P30CA023108.

\bibliographystyle{unsrt}

\end{document}